\documentclass[final,5p,times,twocolumn]{elsarticle}
\usepackage{amssymb}
\usepackage{amsmath}
\usepackage{xcolor}
\usepackage{siunitx}
\DeclareSIUnit\curie{Ci}
\DeclareSIUnit\ucurie{\mu Ci}
\DeclareSIUnit\Mbecquerel{MBq}

\usepackage{hyperref}
\usepackage{subcaption}
\usepackage{lineno}

\journal{Nuclear Instruments and Methods in Research A}
\let\oldbibliography\thebibliography
\renewcommand{\thebibliography}[1]{%
  \oldbibliography{#1}%
  \setlength{\itemsep}{0pt}%
}

\begin{document}

\begin{frontmatter}

\title{Optimization of Cryogenic Detector Test Station by Rejecting Electromagnetic Interference}


\author[PHY]{Sangbaek Lee\fnref{fn2}\corref{cor1}}
\cortext[cor1]{Corresponding author}
\ead{sangbaek.lee@temple.edu}
\author[PHY]{Whitney Armstrong}
\author[MSD]{Maximo DiPreta}
\author[PHY]{Jacob Dulya\fnref{fn3}}
\author[MSD]{Valentine Novosad}
\author[PHY]{Tomas Polakovic}
\affiliation[PHY]{organization={Physics Division, Argonne National Laboratory},%Department and Organization
            %addressline={9700 S. Cass Ave}, 
            city={Lemont},
            state={IL 60439},
            country={U.S.A.}}
\affiliation[MSD]{organization={Materials Science Division, Argonne National Laboratory},%Department and Organization
            %addressline={9700 S. Cass Ave}, 
            city={Lemont},
            state={IL 60439},
            country={U.S.A.}}
\fntext[fn2]{Present address: Temple University}
\fntext[fn3]{Present address: Illinois Institute of Technology}

\begin{abstract}
We report on the solution optimized for characterizing SNSPDs by rejecting electromagnetic interference (EMI) from various sources. The proposed readout method enhances measurement stability and enables reliable device characterization at low bias currents, where the signal-to-noise ratio is typically limited. By effectively suppressing EMI-induced noise, the method improves the ability to distinguish genuine detection events from spurious signals and reduces the effort required for data analysis. The approach has been applied to preliminary measurements of SNSPDs exposed to $\alpha$ particles emitted from a $^{241}$Am source, demonstrating stable operation and clean signal acquisition. While a detailed study of $\alpha$ detection is underway, the method establishes a foundation for further characterization of SNSPDs with various incident particles. The demonstrated EMI rejection technique is expected to facilitate future research in particle detection and support ongoing SNSPD development for applications in nuclear and accelerator-based experiments.
\end{abstract}


% \begin{highlights}
% \end{highlights}


\end{frontmatter}
% \linenumbers %uncomment this after arXiv submission

\section{Introduction}
\label{sec:intro}

%what is the SNSPD?
Superconducting nanowire single-photon detectors (SNSPDs) are state-of-the-art devices designed to detect individual photons across a broad wavelength range, from the near-infrared to the X-ray region. Their operation relies on an abrupt change in resistivity that occurs when a photon is absorbed by the superconducting nanowire \cite{doi:10.1063/1.1388868}. SNSPDs are distinguished by their exceptional timing resolution \cite{Korzh:2018oqv} and high detection efficiency \cite{Marsili:2012oad}, making them highly suitable for a wide range of advanced sensing and measurement applications \cite{nano10061198}.

%What it is good for?
While SNSPDs were originally developed to detect photons in the visible and near-infrared wavelength ranges, their capabilities extend well beyond these traditional applications. SNSPDs can also detect charged particles through two distinct mechanisms: one analogous to photon detection and the other governed by thermal processes \cite{sherman1962superconducting}. The detection of charged particles has been demonstrated across a wide range of energies, including the keV \cite{zen2009development, casaburi20122, ohkubo2008superconducting, suzuki2008ultrafast, sano2015demonstration, suzuki2011hot, doi:10.1126/sciadv.adj2801}, MeV \cite{azzouz2012efficient}, and GeV regimes \cite{Lee:2023brm}. The capability of SNSPDs to detect high-energy charged particles has drawn increasing interest for potential applications in future collider experiments, such as the electron-ion collider (EIC) \cite{Lee:2023brm} and the future circular collider (FCC) \cite{Pena:2024etu}.

%What we are going to report?
SNSPDs are typically characterized inside a cryostat \cite{10.1093/acprof:oso/9780198570547.001.0001}, where EMI from various sources can contaminate the signal channels.\ While radio-frequency (RF) noise can often be mitigated using a Faraday cage or by the metallic enclosure of the cryostat itself, interference originating from three-phase motor components requires a careful analysis of the conductive paths \cite{4957755}.
In this paper, we present a simple and effective method to reliably characterize cryogenic detectors by rejecting such interference using a reference nanowire as an antenna.
This approach provides a practical foundation for building future SNSPD test systems. Characterization measurements have been performed at Argonne National Laboratory (ANL) using an alpha-particle source, which deposits significantly more energy in the nanowire than a 120 GeV proton.
A few-MeV $\alpha$ particle deposits its entire kinetic energy in the silicon substrate, where it is fully stopped, while the 120~GeV proton deposits approximately \SI{0.1}{\MeV} in the substrate \cite{nist_star}.
The deposited energy forms a non-superconducting core of size proportional to the square root of the deposited energy \cite{nano10061198, sherman1962superconducting, brooks1956nuclear, kelsch1962observation}.
The previous work with a 120~GeV proton determined the hot spot size to be 134~nm \cite{Lee:2023brm}, and a 5.5~MeV $\alpha$ particle creates an approximately \SI{1}{\um} hot spot by the same scaling.
In addition to demonstrating the EMI rejection method, this work discusses the SNSPD count rate observed with $\alpha$ particles, providing a valuable opportunity to investigate the ion detection mechanism and to optimize sensor designs for future collider experiments.

\begin{figure}[!ht]
    \centering
    \includegraphics[width=0.8\linewidth]{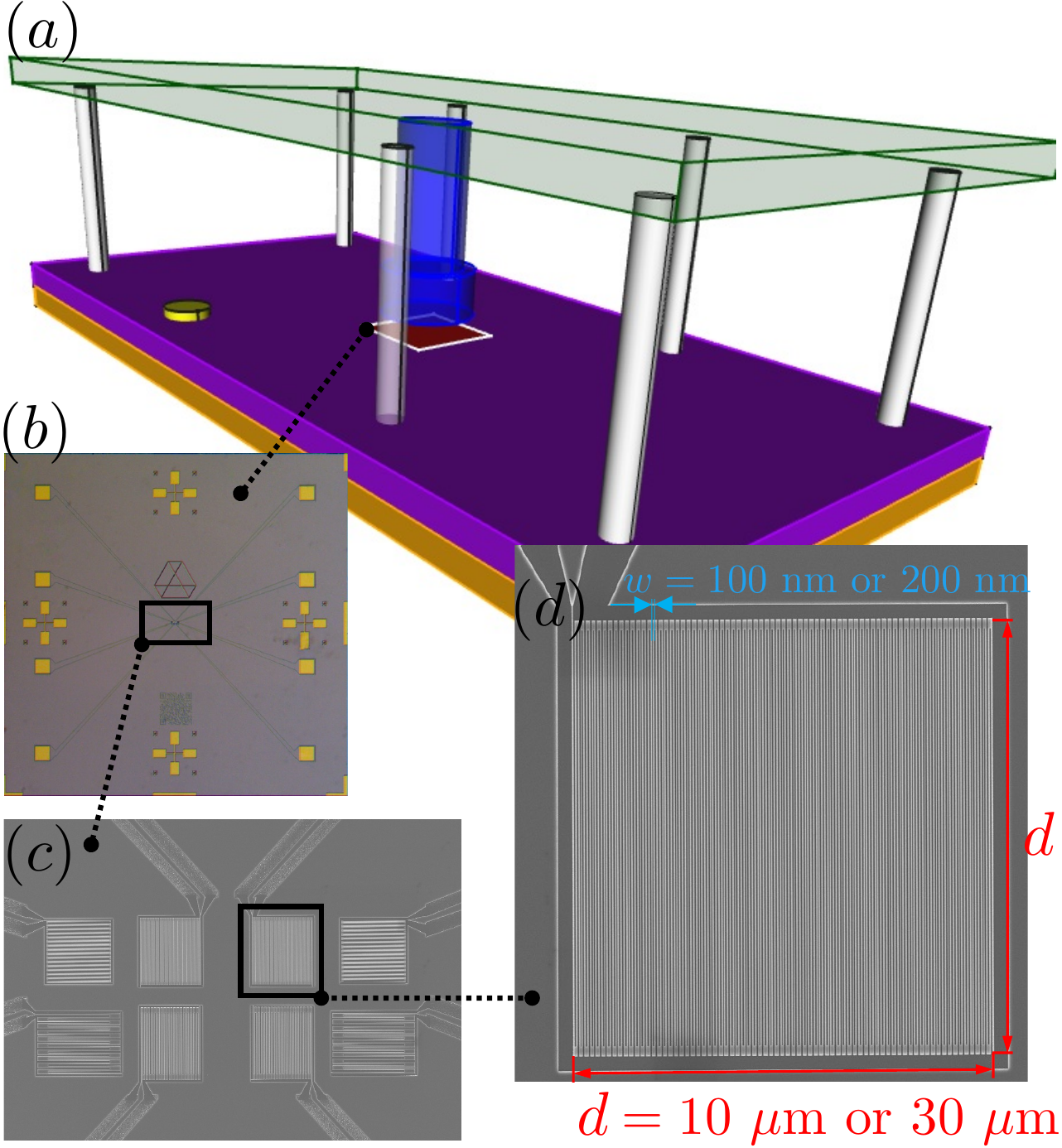}
    \caption{(a) A schematic drawing of the testbed inside the cryostat (not-to-scale). The nanowire chip (red) wire-bonded to the PCB (purple) is detecting the $\alpha$ particles from $^{241}$Am source (blue). The PCB support (orange) is made of copper and attached to the second stage of cryostat, temperature of which is monitored by the silicon diode sensor (yellow). The sealed source support (green) is fixed to the PCB support by standoffs. (b) A zoom-in optical microscope image of the nanowire chip. Eight nanowire devices are patterned on the center of the chip, with wire-bonding pads visible at the sides. (c) A SEM image of the eight nanowire devices. One nanowire device is bordered by a black rectangle. (d) A zoom-in SEM image of the nanowire device. A meandering nanowire of width $w$ (light blue) is patterned in a square of side length $d$ (red). Dotted lines with filled circles indicate the zoom-in regions connecting successive panels.}
    \label{fig:schematic}
\end{figure}
\section{Experimental Setup}
\label{sec:experiment}

Figure~\ref{fig:schematic}(a) shows a schematic of the testbed inside the cryostat used to characterize $\alpha$-particle detection with SNSPDs. Nanowire chips with the nanowire sensors of different geometries were mounted inside a cryostat coupled to a Gifford-McMahon (GM) cryocooler. The fabrication of these devices has been described in previous studies \cite{ionBeamNbN, Lee:2023brm}. Two nanowire chips were used in this study. An optical microscope image of a nanowire chip is shown in Fig.~\ref{fig:schematic}(b). Each chip has 8 nanowire devices of width $w$ patterned in a square active area of side length $d$ as seen in Fig.~\ref{fig:schematic}(c). Each device consists of a single continuous superconducting nanowire patterned in a meander geometry as shown in Fig.~\ref{fig:schematic}(d). Devices of Chip~1 were fabricated with $w \in \{100, 200, 300, 400\}$~nm and $d=\SI{10}{\um}$, and those of Chip~2 with $w \in \{100, 200, 400, 800\}$~nm and $d=\SI{30}{\um}$. Each chip was wire-bonded to a PCB equipped with 50~$\Omega$ shunt resistors. Three geometries of the device under testing (DUT) were studied in this work: (a) $w=\SI{100}{\nm}$ from Chip~1, (b) $w=\SI{200}{\nm}$ from Chip~1, and (c) $w=\SI{200}{\nm}$ from Chip~2. The PCB was mounted on a copper tee attached to the second stage of the GM cryocooler, maintaining a cryogenic temperature of $T=\SI{2.7}{K}$. The temperature at the PCB was monitored using a silicon diode sensor.

Among possible superconducting nanowire material choices, Niobium has an advantage in its radiation hardness \cite{10.1063/1.341850, Smith1973, PhysRevB.79.094509, Piatti2016} in the relatively higher operating temperature of Nb-compounds such as NbN. No radiation damage or degradation of detection efficiency of SNSPDs was reported after irradiation over a few days \cite{azzouz2012efficient}.

\begin{figure}[!ht]
    \centering
    \includegraphics[width=0.8\linewidth]{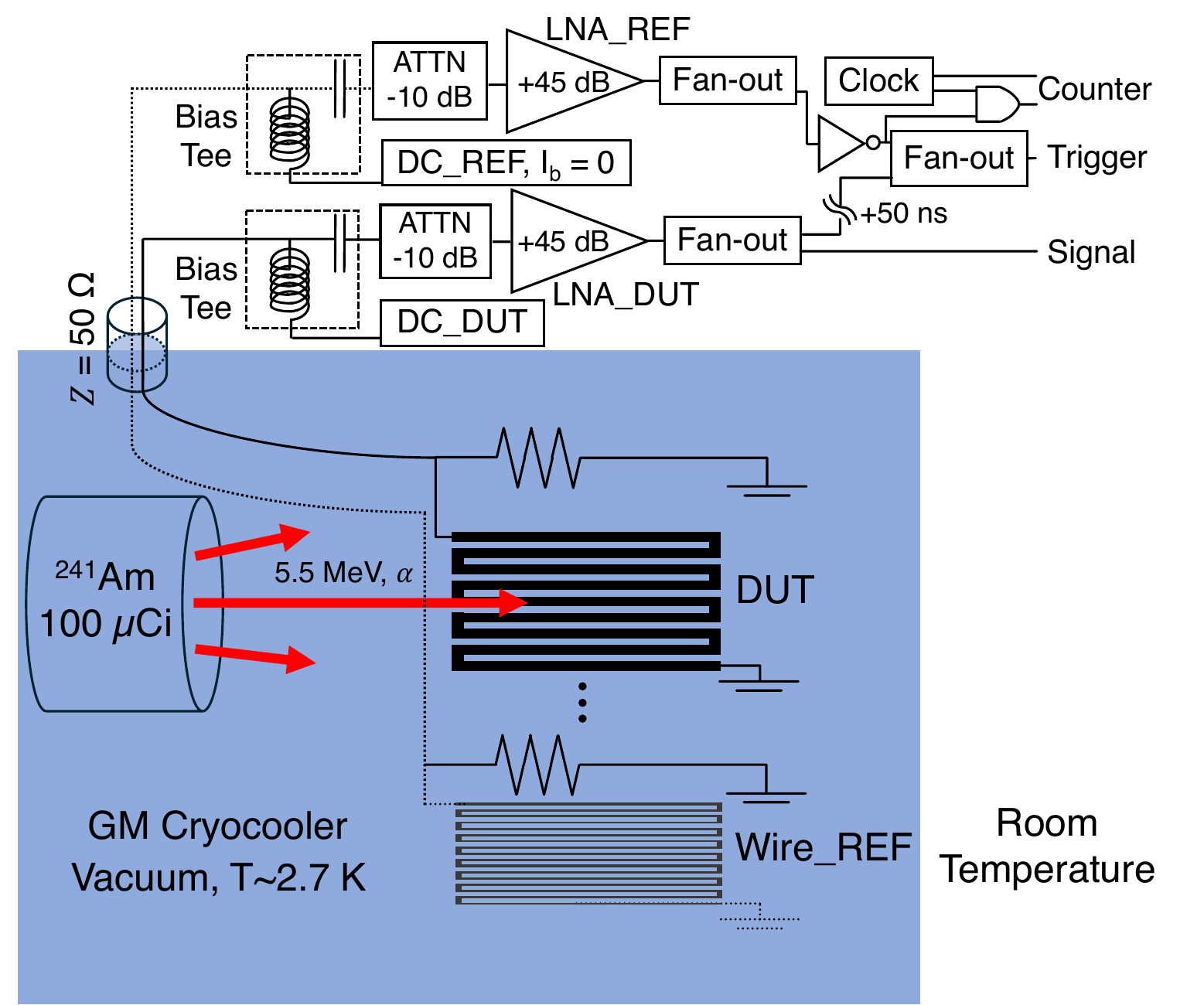}
    \caption{A block diagram depicts the electronic circuitry inside the cryostat (blue background) and outside for the logic-gated mode (white background). The $\alpha$ particle detection was carried out by the DUT. For the logic-gated mode, the raw voltage waveform of signal and interference was monitored on the oscilloscope, with trigger logic implemented to veto interference signals.}
    \label{fig:block_diagram}
\end{figure}

The nanowire devices were tested both with and without a $^{241}$Am source of \SI{100}{\ucurie} (equivalent to \SI{3.7}{\Mbecquerel}) mounted on a source holder  (blue cylinder of Figure~\ref{fig:schematic}(a)). The source was mounted on metallic, thermally conductive aluminum standoffs, ensuring it was thermalized to the same cryogenic temperature of $T=\SI{2.7}{\K}$ as the PCB, which is required for superconducting operation of the nanowire devices. Since radioactive decay is temperature-independent, this thermalization does not affect the $^{241}$Am activity. This source emits $\alpha$ particles with energies of \SI{5.49}{\MeV} (84.8\%), \SI{5.44}{\MeV} (13.1\%), and \SI{5.34}{\MeV} (1.7\%), as well as significant hard X-rays at \SI{59.54}{\keV} (35.9\%) and \SI{26.34}{\keV} (2.3\%) \cite{nndc}. Owing to the long penetration depth of these hard X-rays \cite{nist_database}, a nanowire layer thickness on the order of \SI{100}{\nm} is typically required for X-ray sensitivity \cite{10.1093/nsr/nwad102, 10.1063/1.4759046}. The nanowire device used in this study was \SI{12}{\nm} thick, rendering it effectively insensitive to X-rays. To confirm this, an additional measurement was conducted with an $\alpha$-shield, in which a \SI{0.5}{\mm}-thick stainless-steel layer was placed between the source and the sensor to block the $\alpha$ flux and isolate the response to X-rays. The distance between the radioactive source and the nanowire device $z$ was \SI{4.85}{\mm}, determined from the measured lengths of the standoffs and the source container. While the radius of the source container was \SI{6.35}{\mm}, the source activity was concentrated in the center. The chip was uniformly irradiated by the source, as the source-to-chip distance is large compared to the chip's active area (see Fig.~\ref{fig:schematic}(d)). Given the long half-life of $^{241}$Am (432.6 years) \cite{nndc}, the $\alpha$ flux is considered stable over time. The geometrical acceptance of the detector is approximately $d^2/\left(4\pi z^2\right)$, which is $O(\SI{0.1}{ppm})$ for the $d=\SI{10}{\um}$ device and $O(\SI{1}{ppm})$ for the $d=\SI{30}{\um}$ device. With a source activity of \SI{100}{\ucurie}, the expected count rate is $O(\SI{1}{\Hz})$ for the $d=\SI{10}{\um}$ device and $O(\SI{10}{\Hz})$ for the $d=\SI{30}{\um}$ device.

\begin{figure}[!ht]
    \centering
    \includegraphics[width=0.95\linewidth]{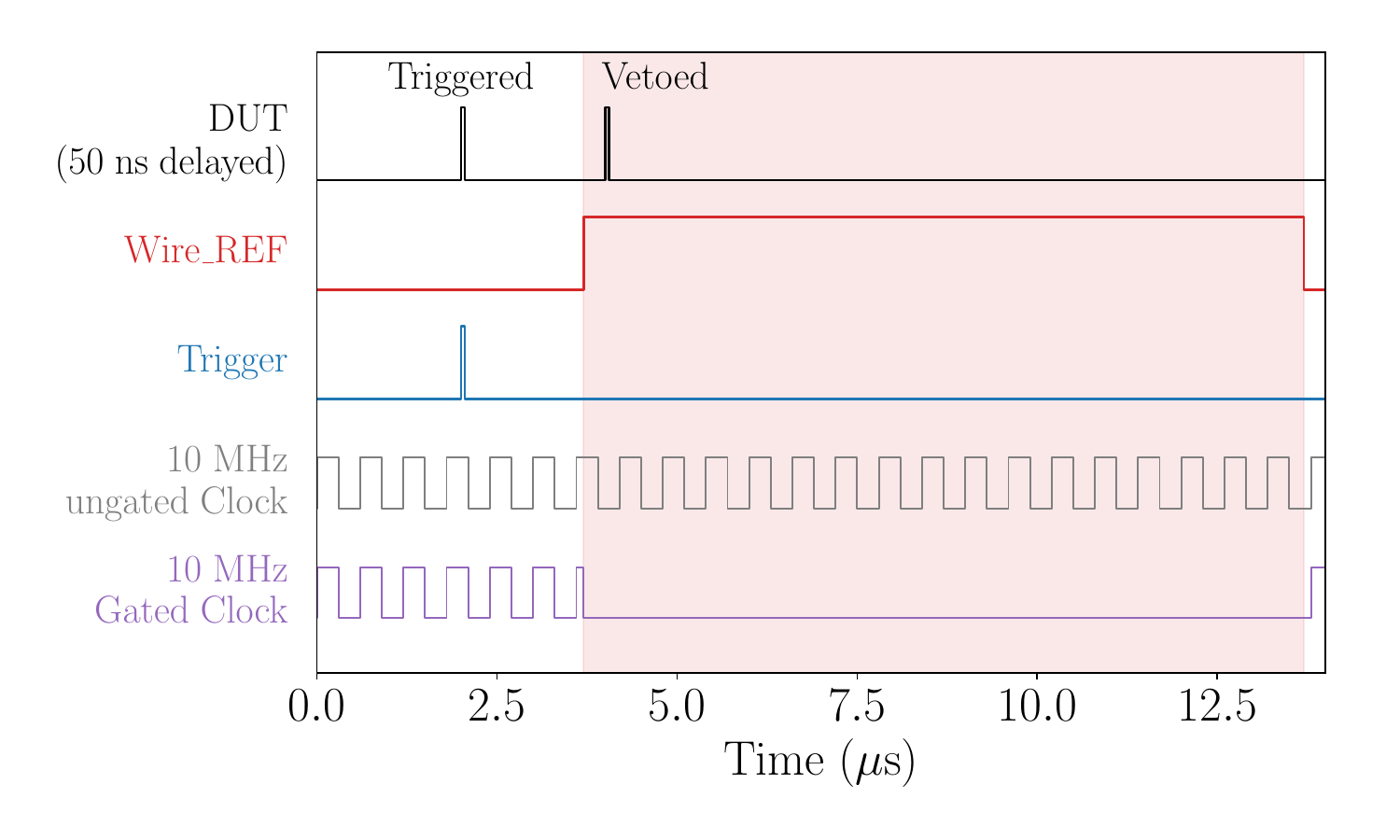}
    \caption{Schematic illustration of the anti-coincidence trigger logic, shown for a representative triggered event without EMI and a vetoed event with EMI. The DUT discriminator output is delayed by 50~ns to align with the veto decision. When Wire\_REF fires in the presence of EMI, it opens a \SI{10}{\micro\second} veto gate that inhibits the trigger, so that the final trigger (DUT AND NOT Wire\_REF) is formed only in the absence of EMI. The ungated and gated 10~MHz clocks are used to determine the system livetime, defined as the ratio of gated to ungated clock counts.}
    \label{fig:timing_diagram}
\end{figure}

Figure~\ref{fig:block_diagram} presents the block diagrams for electronics setup. The DUT was current-biased by an LTC1427 current output DAC (DC\_DUT), with slow control managed by a Raspberry Pi 3 Model B, and its signal was amplified by a low noise amplifier (LNA\_DUT). Two data acquisition modes were used in this work. In the self-triggered mode, only the DUT was recorded by an oscilloscope triggered at a threshold of \SI{100}{\milli\volt}. In the logic-gated mode, both the DUT and the reference nanowire (Wire\_REF) contributed to the trigger. Wire\_REF is a fully functional SNSPD device with the same detector characteristics as the DUT on the same chip and shares the ground with the DUT. It was deliberately operated at zero bias current via a separate DAC (DC\_REF) so that it functions as a passive antenna sensitive to EMI pickup rather than as a detector. A width of $w=\SI{100}{\nm}$ was chosen for Wire\_REF because its response to EMI-induced transients is comparable to or more sensitive than that of the DUT across all geometries studied, ensuring that Wire\_REF reliably fires whenever the DUT is affected by EMI. Figure~\ref{fig:block_diagram} presents the block diagram of the electronics setup for the logic-gated mode. One fan-out output of the DUT signal was recorded by an oscilloscope, while the other was passed through a discriminator with a \SI{40}{\milli\volt} threshold and delayed by 50~ns before being sent to one input of a LeCroy 365A 4-fold logic unit. The EMI around the metal components of the cryostat, primarily arising from the cold head motor and complex return paths, typically manifested as a short pulse on other wires. The interference detected through Wire\_REF (Fig.~\ref{fig:block_diagram}) was amplified by a separate low noise amplifier (LNA\_REF). The Wire\_REF signal was discriminated at the same \SI{40}{\milli\volt} threshold and stretched to a \SI{10}{\micro\second} gate by a dual gate generator, with the gate width chosen to match the typical duration of the EMI pulses, before being sent to the second input of the LeCroy 365A as a veto. The 365A unit then produces an AND-NOT trigger, accepting the 50-ns delayed DUT signal only when no Wire\_REF veto gate is active, thereby rejecting EMI-correlated triggers while preserving genuine $\alpha$-induced events on the DUT. The trigger thresholds for both modes were determined by the noise floor of the electronics. For the self-triggered mode, the oscilloscope threshold was set to \SI{100}{\milli\volt} for detectors (a) and (b), and \SI{150}{\milli\volt} for detector (c). For the logic-gated mode, a discriminator threshold of \SI{40}{\milli\volt} was used for all three detectors.

\begin{figure}[!t]
    \centering
    \includegraphics[width=0.8\linewidth]{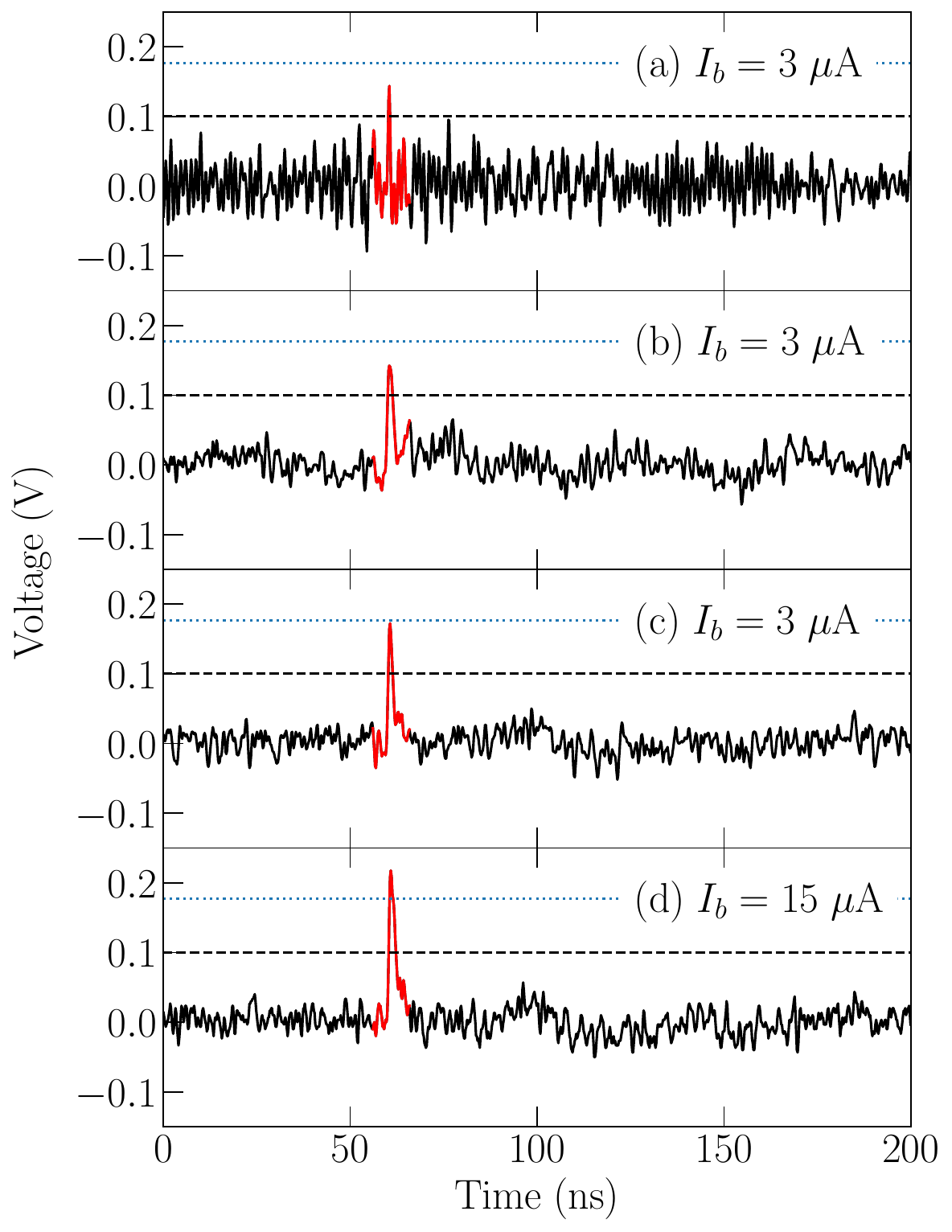}
    \caption{Representative recorded waveforms in the self-triggered mode (black) with the candidate pulse highlighted in red. The black dashed line indicates the \SI{100}{\milli\volt} threshold used in the self-triggered mode, and the blue dotted line indicates the highest EMI-induced pulse amplitude observed, \SI{177}{\milli\volt}. Panels (a)--(c) were recorded at $I_b=\SI{3}{\micro\ampere}$, a bias current too low to produce a genuine $\alpha$-induced pulse on the DUT, so the pulses shown are attributable to EMI rather than a true detector response. The three panels illustrate a progression from baseline noise (a) to an EMI-induced pulse of intermediate amplitude (b) to one approaching the highest observed EMI amplitude (c). Panel (d), recorded at $I_b=\SI{15}{\micro\ampere}$, shows a pulse plausibly attributable to a genuine $\alpha$-induced event. The waveform shapes in (c) and (d) are not clearly distinguishable, indicating that amplitude and shape alone are insufficient to discriminate EMI-induced pulses from genuine detector pulses at comparable amplitude.}
    \label{fig:waveforms}
\end{figure}

The DAQ livetime was estimated by counting the cycles of a \SI{10}{\mega\hertz} external clock using a LeCroy 2551 Scaler in conjunction with a CAEN C111C Ethernet CAMAC Crate Controller. Two clock signals were used: an ungated clock (Clock1) running continuously, and a gated clock (Clock2) inhibited during the Wire\_REF veto window. A livetime of 99.5\% was measured from the ratio of Clock2 to Clock1 counts, confirming that the chosen \SI{10}{\micro\second} gate width introduces negligible deadtime while effectively suppressing EMI-induced triggers. Figure~\ref{fig:timing_diagram} illustrates this anti-coincidence trigger logic schematically, showing both a triggered event in the absence of EMI and a vetoed event when EMI is present. The effect of logic was studied by taking the reference data without using logic components, i.e., by taking the data with self-triggered LNA\_DUT output using only the DUT. In the logic-gated mode, both the DUT and Wire\_REF contribute to the trigger: the DUT signal forms the candidate trigger, while Wire\_REF serves exclusively as the veto channel via the AND-NOT logic described above.

\section{Results}
\label{sec:results}

Figure~\ref{fig:waveforms} shows representative recorded waveforms for the device with $w=\SI{200}{\nm}$ and $d=\SI{10}{\um}$ in the self-triggered mode. Panels (a)--(c) were recorded at $I_b=\SI{3}{\micro\ampere}$, a bias current too low to produce a genuine $\alpha$-induced pulse on the DUT. Any pulse exceeding the threshold should be attributed to EMI-induced signal rather than to a true detector response. These panels illustrate examples of (a) a baseline noise waveform, (b) an intermediate waveform showing signal in the presence of noise, and (c) a cleaner signal waveform without a clear fingerprint of noise. Panel (d), recorded at $I_b=\SI{15}{\micro\ampere}$, shows a pulse plausibly attributable to a genuine $\alpha$-induced signal. The waveform shapes in Figs.~\ref{fig:waveforms}(c) and~(d) are not clearly distinguishable, indicating that amplitude and shape alone are insufficient to discriminate EMI-induced pulses from genuine detector pulses at comparable amplitude. This motivates the coincidence-based veto approach described in this work rather than relying on pulse-shape discrimination.

\begin{figure}[!ht]
    \centering
    \includegraphics[width=0.8\linewidth]{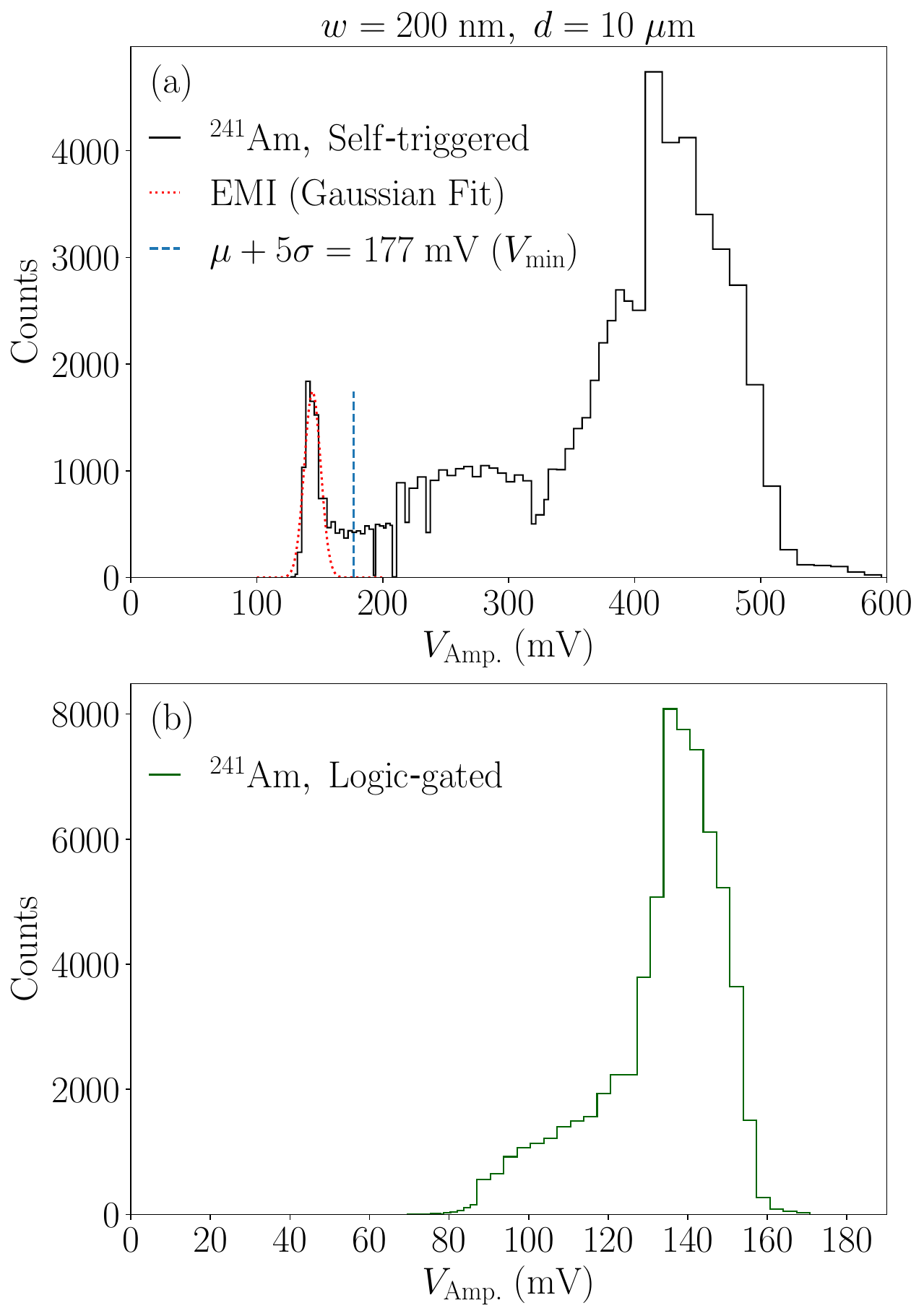}
    \caption{(a) A histogram of $V_{\mathrm{amp}}$ accumulated over all bias current settings used in the self-triggered mode, for the device with \textit{w} = \SI{200}{\nm} and \textit{d} = \SI{10}{\um} (black). The peak around \SI{150}{\milli\volt} indicates substantial EMI contributions measured alongside genuine detector signals. This peak is fitted with a Gaussian profile (red dotted), and $\mu + 5\sigma$ is adopted as the minimum voltage-amplitude threshold (blue dashed) for Fig.~\ref{fig:results}, where $\mu$ and $\sigma$ denote the mean and standard deviation of the Gaussian fit. (b) The corresponding $V_{\mathrm{amp}}$ histogram in the logic-gated mode, where the population of high-amplitude EMI-induced events is strongly suppressed (green). As described in Sec.~\ref{sec:experiment}, the two datasets were recorded using different acquisition modes (self-triggered versus logic-gated) that have different signal bandwidth. The data collection durations for each bias current were not designed to be identical for the two measurements. For these reasons, panels (a) and (b) are shown separately and should be interpreted as a qualitative illustration of EMI suppression rather than a quantitative comparison.}
    \label{fig:vamp_cut}
\end{figure}

Figure~\ref{fig:vamp_cut} shows the voltage amplitude ($V_{\mathrm{amp}}$) histograms for the device with $w=\SI{200}{\nm}$ and $d=\SI{10}{\um}$. Panel~(a), recorded in the self-triggered mode, shows a Gaussian peak around \SI{150}{\milli\volt} indicating substantial EMI contributions, and demonstrates that measurements without EMI rejection require a minimum voltage-amplitude threshold defined as $\mu + 5\sigma$ above the mean of the Gaussian fit. Panel~(b), recorded in the logic-gated mode, shows the same device with EMI rejection enabled, where the population of high-amplitude EMI-induced events is strongly suppressed.

\begin{figure}[!ht]
    \centering
    % Uncomment this for the peer review
    \includegraphics[width=.8\linewidth]{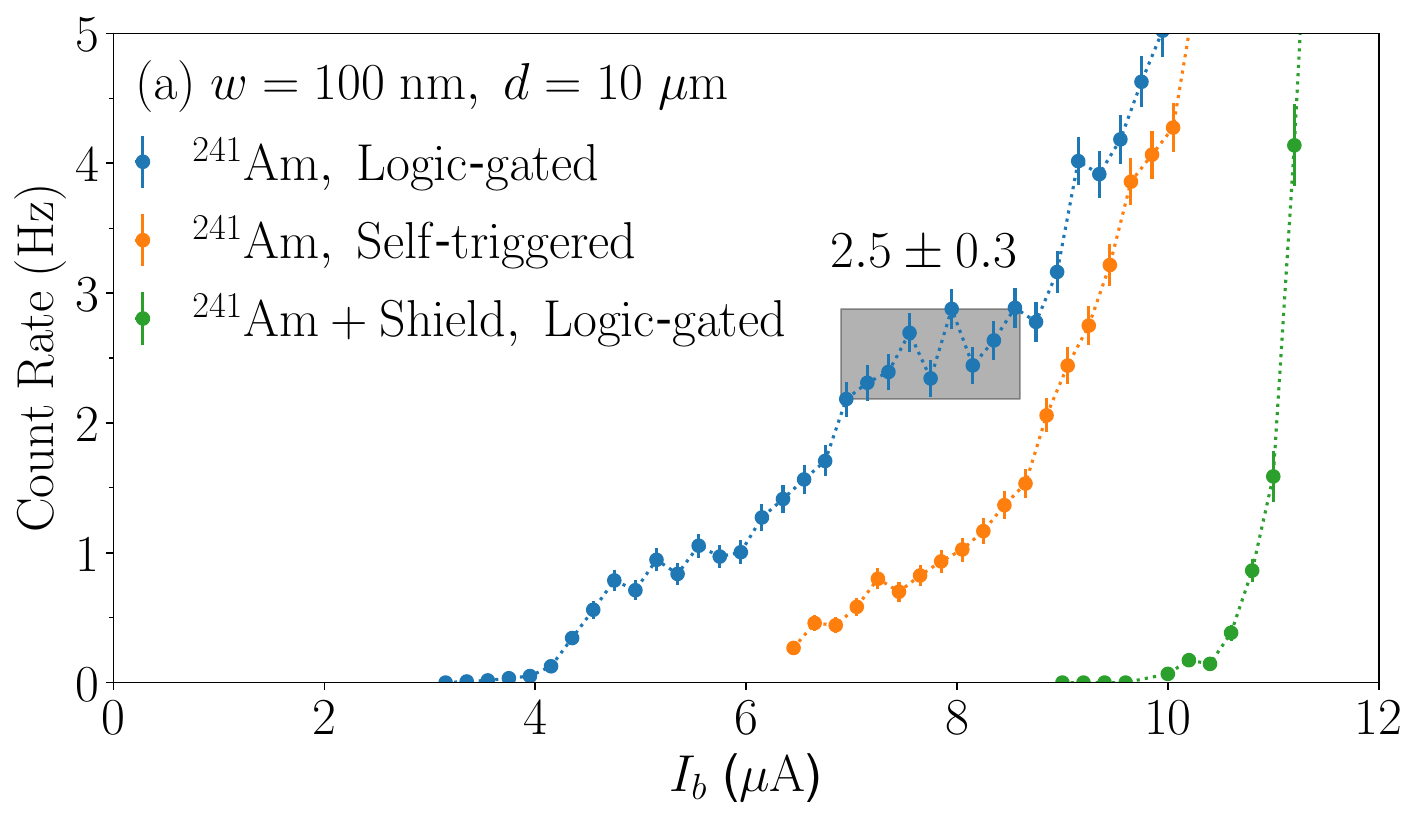}\\
    \includegraphics[width=.8\linewidth]{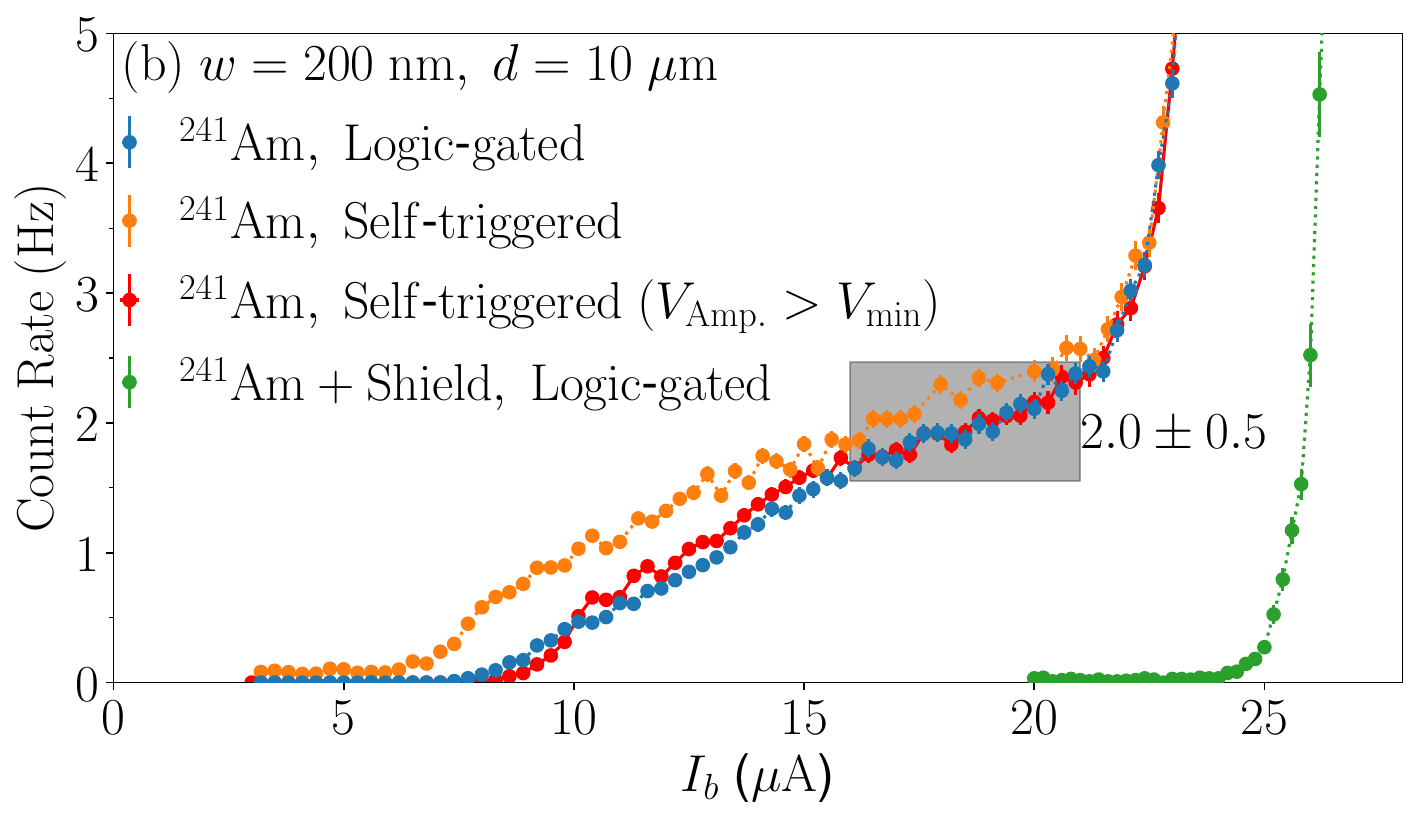}\\
    \includegraphics[width=.8\linewidth]{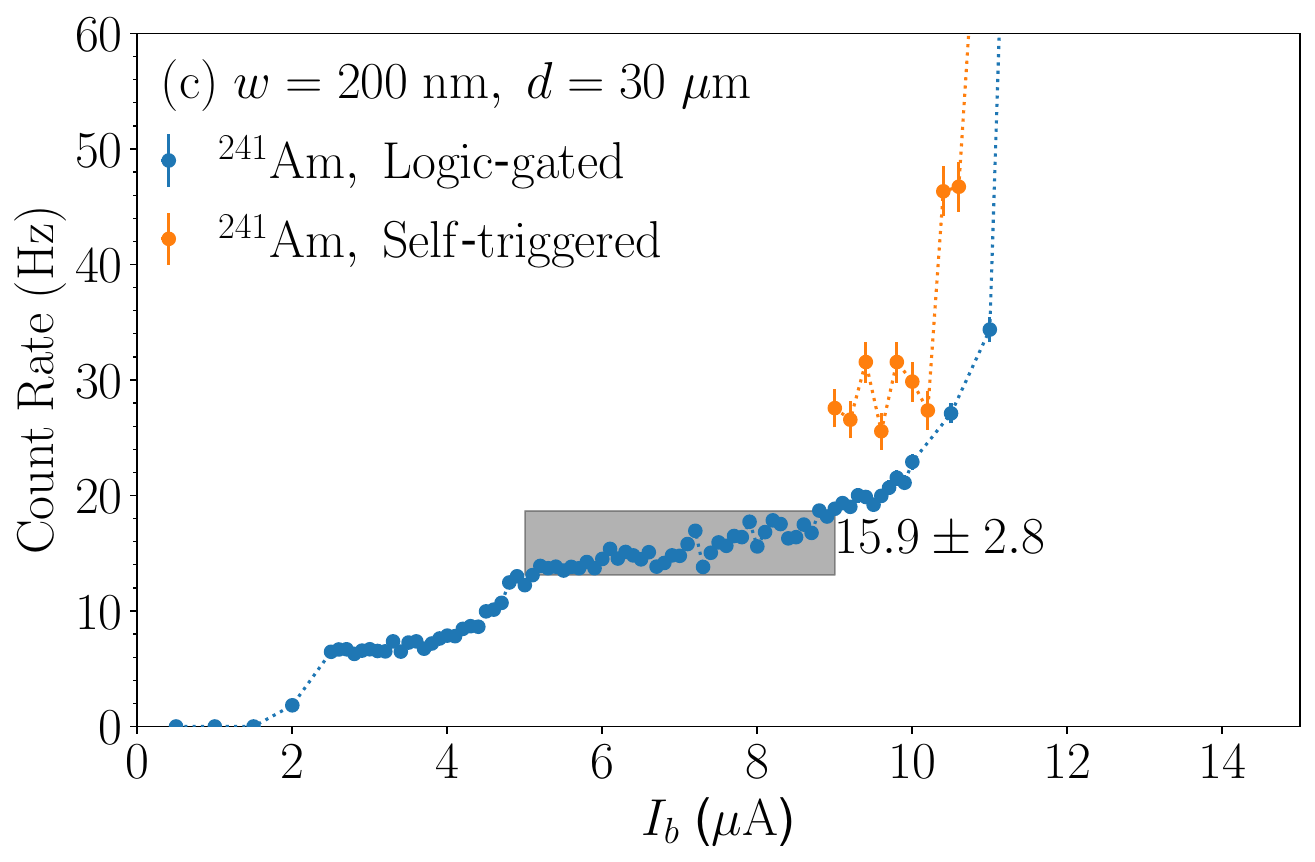}
    % Comment out this after the arXiv submission.
    % \includegraphics[width=\linewidth]{fig3_100nm.pdf}
    % \includegraphics[width=\linewidth]{fig3_200nm.pdf}
    % \includegraphics[width=\linewidth]{fig3_200nm_30um.pdf}
    \caption{Measured count rates as a function of bias current for SNSPDs with varying geometries. Data are shown with EMI rejection (blue), without EMI rejection (orange), and with an additional $\alpha$-shield (green) where applicable. The gray rectangles and adjacent numbers indicate the estimated count rate at the plateau. (a) \SI{100}{\nm}-wide, \SI{10}{\um}-long device: EMI suppression is essential at low bias to reveal $\alpha$-induced counts above the noise floor. (b) \SI{200}{\nm}-wide, \SI{10}{\um}-long device: EMI rejection removes spurious interference counts. For this case, rates measured with the minimum voltage amplitude condition defined in Fig.~\ref{fig:vamp_cut}(a) (red) agree with the EMI-rejected data (blue). (c) \SI{200}{\nm}-wide, \SI{30}{\um}-long device: EMI rejection is crucial for obtaining stable and reliable measurements of the SNSPD response.}
    \label{fig:results}
\end{figure}

Figure~\ref{fig:results} shows the measured count rate as a function of bias current for three SNSPD devices with different geometries. In all panels, data are compared between measurements performed with and without EMI rejection, and with an additional $\alpha$-shield when applicable.

(a) For the device with \textit{w} = \SI{100}{\nm} and \textit{d} = \SI{10}{\um}, the noise floor is significantly elevated in the self-triggered mode. The higher oscilloscope threshold required in this case rejects genuine low-amplitude $\alpha$-induced pulses, resulting in a lower observed count rate (orange curve in Fig.~\ref{fig:results}(a)) than the logic-gated mode (blue curve). Therefore, effective EMI suppression is crucial for achieving reliable results in this regime, particularly at lower bias currents where the onset of $\alpha$-induced counts is characterized.

(b) For the device with \textit{w} = \SI{200}{\nm} and \textit{d} = \SI{10}{\um}, operating above a bias current of \SI{8}{\micro\ampere} enables measurements in the self-triggered mode above the noise floor across the entire bias range. In the absence of EMI rejection, a substantial number of noise-induced events are observed, as evidenced by a nonzero count rate even at low bias currents (orange curve in Fig.~\ref{fig:results}(b)). Implementing EMI rejection effectively suppresses these spurious counts (blue curve). Applying the threshold of \SI{177}{\milli\volt} defined in Fig.~\ref{fig:vamp_cut}(a) yields a count rate (red curve) that agrees well with the EMI-rejected dataset (blue curve). This confirms that EMI rejection suppresses noise-induced events without affecting genuine detector responses. Moreover, the EMI rejection method does not require fitting $V_{\mathrm{amp}}$ or recording individual waveforms, enabling more efficient data acquisition.

For the devices used in measurements (a) and (b), similar measurements were also performed with a \SI{0.5}{\mm}-thick stainless steel shield as shown by green curves in Fig.~\ref{fig:results}(a)--(b). The application of the shield significantly reduces the observed counts, demonstrating that the detector response does not originate from $\gamma$ rays emitted by $^{241}$Am.

(c) For the larger-area device ($w=\SI{200}{\nm}$, $d=\SI{30}{\um}$), a reliable measurement of count rate in the self-triggered mode was not feasible due to the elevated noise level. Chip~2 hosting this device was fabricated from a different NbN film than Chip~1, so the switching current differs from that of geometry (b) despite the identical wire width. Because the device is wire-bonded to a different PCB, the characterization measurements were performed in a different cooldown period from those of (a) and (b). During this cooldown period, the EMI-induced noise was elevated, possibly due to differences in the cabling path, and the oscilloscope threshold was set to \SI{150}{\milli\volt}. Similarly to (b), in the absence of EMI rejection, an excess count rate above $I_b=\SI{8}{\micro\ampere}$ is observed in the self-triggered mode (orange curve), which can be attributed to EMI-induced events. In the logic-gated mode, the lower \SI{40}{\milli\volt} threshold permits a genuine $\alpha$-induced plateau to be observed over a bias-current range where the noise rate remains subdominant, exhibiting a clean transition in count rate near the switching current, confirming stable detector operation and demonstrating the importance of interference suppression for large-area SNSPDs.

The count rate at the plateau is measured to be 2.5~$\pm$~0.3~Hz for the $w=\SI{100}{\nm}$, $d=\SI{10}{\um}$ device, 2.0~$\pm$~0.5~Hz for the $w=\SI{200}{\nm}$, $d=\SI{10}{\um}$ device, and 15.9~$\pm$~2.8~Hz for the $w=\SI{200}{\nm}$, $d=\SI{30}{\um}$ device (gray rectangles in Fig.~\ref{fig:results}). The observed plateau count rates are consistent with the expected rates (Sec.~\ref{sec:experiment}) to within the same order of magnitude. A precise determination of absolute detection efficiency is beyond the scope of this work. The small discrepancy is within the systematic uncertainties associated with the source geometry, hot spot size, potential fabrication variations in the NbN film, and the simplified hot spot model itself, none of which have been rigorously characterized here. The count-rate ratio between the $d=\SI{30}{\um}$ and $d=\SI{10}{\um}$ devices is 8.0~$\pm$~2.4, consistent with the naive geometric factor of (\SI{30}{\um})$^2$/(\SI{10}{\um})$^2=9$ from the ratio of active areas. These results are consistent with expectations that the energy deposited by $\alpha$ particles is sufficient for detection in the \SI{100}{\nm} wire, which is narrower than the \SI{250}{\nm} devices optimal for 120~GeV proton detection \cite{Lee:2023brm}.
\section{Summary}
\label{sec:summary}

We presented a readout method designed to reject EMI by using additional SNSPD on the same chip as an antenna, and thereby improve the signal fidelity in SNSPD measurements. This method enables reliable device characterization even at low bias currents, where the signal-to-noise ratio is typically limited. In addition, the approach allows researchers in SNSPD R\&D to efficiently distinguish genuine signal waveforms from noise, saving both time and computational resources during data analysis.

While a detailed analysis of $\alpha$-particle detection with SNSPDs is beyond the scope of this paper, our results demonstrate that SNSPDs can effectively register $\alpha$ events. The ability to detect $\alpha$ particles has important implications for a range of applications, including nuclear physics experiments, radiation monitoring, and particle identification studies.

Furthermore, the demonstrated EMI rejection and stable $\alpha$ detection performance highlight the versatility of SNSPDs as particle sensors beyond their traditional use as photon detectors. We anticipate that the methods and findings presented here will support future research in SNSPD-based particle detection and contribute to the development of high-precision, low-noise detection systems for both fundamental studies and applied technologies.
% \appendix

 % \bibliographystyle{elsarticle-num} 
 % \bibliography{cas-refs}





% \end{thebibliography}

\section*{Acknowledgements}
%We would like to thank Fermilab Test Beam Facility staff and engineers for their technical support and infrastructure accommodations essential for these measurements.
%  This material is based upon work supported by the U.S. Department of Energy, 
%  Office of Science, Office of Nuclear Physics, under contract number 
%  DE-AC02-06CH11357.
%
% Isn't the contract number below specific to NP? (Not BES/MSD?) If so, we need to fix this.
%  
This material is based upon work supported by the {U.S. Department of Energy, Office of Science, Office of Nuclear Physics}, under contract number {DE-AC02-06CH11357}; by the {U.S. Department of Energy, the Office of Science} under the Microelectronics Co-Design Research Project ``Hybrid Cryogenic Detector Architectures for Sensing and Edge Computing enabled by new Fabrication Processes'' ({LAB 21-2491}); and by the {U.S. Department of Energy, Office of Science, Offices of High Energy Physics and of Basic Energy Sciences}, as part of the Co-design and Heterogeneous Integration in Microelectronics for Extreme Environments (CHIME) Microelectronics Science Research Center (MSRC), under contract number {LAB 24-3320}. W.~A. and T.~P. acknowledge financial support from the {U.S. Department of Energy, Office of Nuclear Physics}, Early Career Award under grant number~{DE-FOA-0003176}.

%This work was supported by the U. S. Department of Energy (DOE), Office of 
%Science, Offices of Nuclear Physics,
%Basic Energy Sciences, Materials Sciences and Engineering Division under 
%Contract \# DE-AC02-06CH11357.
\bibliography{references}
\bibliographystyle{elsarticle-num}

\end{document}